\documentstyle[aps]{revtex}
\begin{document}
\draft
\title{Glauber Critical Dynamics: Exact Solution of the Kinetic Gaussian Model}
\author{Jian-Yang Zhu$^{a,b,c}$\thanks{%
E-mail: zhujy@bnu.edu.cn} and Z. R. Yang$^{a,b}$}
\address{$^a$CCAST (World Laboratory), Box 8730, Beijing 100080, China\\
$^b$Department of Physics, Beijing Normal University, Beijing 100875, China%
\thanks{%
Mailing address}\\
$^c$Department of Physics, Jiangxi Normal University, Nanchang 330027,China }
\maketitle

\begin{abstract}
In this paper, we have exactly solved Glauber critical dynamics of the
Gaussian model on three dimensions. Of course, it is much easy to apply to
low dimensional case. The key steps are that we generalize the spin change
mechanism from Glauber's single-spin flipping to single-spin transition and
give a normalized version of the transition probability . We have also
investigated the dynamical critical exponent and found surprisingly that the
dynamical critical exponent is highly universal which refer to that for one-
two- and three-dimensions they have same value independent of spatial
dimensionality in contrast to static (equilibrium) critical exponents.
\end{abstract}

\pacs{PACS numbers: 64.60.Ht, 75.10.Hk}

\section{Introduction}

Irreversible dynamic systems exhibit complicated and interesting
non-equilibrium phenomena near the critical point. The study of
non-equilibrium statistical mechanics is much more difficult than
equilibrium state due to the complexity. However, the interesting dynamic
critical behaviors have been attracting a large number of researchers to
work hard for many years.

Up to now, there is no general theory based on the first principle to
describe the dynamic problems. However, a great progress has been made since
the pioneering work completed by Glauber\cite{1} and Kawasaki\cite{2}.
According to their theory, the time-evolving of the order parameters is
described by Markov processes with Glauber single spin flipping mechanism or
Kawasaki exchange mechanism between two spins. Since then people have been
paid much attentions to the study of critical dynamics. The research so far
have been extended from the kinetic Ising model to the kinetic Potts model
and from integer to fraction dimension, in which many approximate methods
such as Monte Carlo simulation, high-temperature series expansion, $\epsilon 
$-expansion, bond-moving renormalization-group method, etc. have been applied%
\cite{3,19}.

Now let us turn on master equation, a basic equation for treating critical
dynamics. As we know, the key step for solving master equation is the
determination of the transition probability. Usually the transition
probability between different states, i.e., different spin-configurations of
the system, is only determined in terms of the detailed balance condition.
Since such a choice is not unique, then leaves behind some arbitrariness.
For removing the arbitrariness, at least in part, we suggest a normalized
transition probability, this means in unit time interval the transition may
occurs or may not. We apply this point to continuous spin (Gaussian) system
and obtain exact solutions of one-, two- and three dimensional kinetic
Gaussian model.

This paper is organized as follows. In Sec.\ref{Sec.2}, a single-spin
transiting critical dynamics which suits arbitrary spin systems is
presented. As an application, the Gaussian model is treated in Sec.\ref
{Sec.3}. We not only obtain the exact solutions of 1D, 2D and 3D kinetic
Gaussian model, but also find that the dynamic critical exponent is highly
universal. Finally, the conclusion is devoted to Sec.\ref{Sec.4}.

\section{Formalization}

\label{Sec.2}

For an irreversible dynamic system subjected to the time-dependent
perturbation, once the perturbation is removed, the system will very slowly
approach to equilibrium state because of the large-scale fluctuation near
the critical point, which is what we call the critical slowing down
phenomenon. The critical dynamics will attempt to explain why the
short-range interactions lead to long-time relaxation. Due to the complexity
there have been no microscopic theory based on the first principle so far,
thus a suitable theoretical model will be quite important. As already
mentioned, both the Glauber dynamics and Kawasaki dynamics have proven to be
successful in many dynamic systems.

In this section, we give a brief review of Glauber dynamics. Of course, we
will give some improvement so that it can suits arbitrary-spin systems. For
clearness, we start from the one-dimensional case (1D), and then the
formulation is easy to extend to 2D and 3D.

The 1D lattice-spin model we shall discuss is a stochastic one. The spins of 
$N$ fixed particles are represented as stochastic functions of time $\sigma
_j(t)$, ($j=1,\cdot \cdot \cdot ,N$), which can be taken discrete values
(discrete-spin model) or continuous values (continuous-spin model), and make
transitions among these values. The transition, according to the Glauber
dynamics, can only change single spin value each time, such as $\sigma
_j(t)\rightarrow \hat{\sigma}_j(t)$ because of the interacting of the system
with heat reservoir. The transition probability $W_j(\sigma _j(t)\rightarrow 
\hat{\sigma}_j(t))$ from configuration $\left( \sigma _1(t),\sigma
_2(t),\cdots ,\sigma _j(t),\cdots ,\sigma _N(t)\right) $ to configuration $%
\left( \sigma _1(t),\sigma _2(t),\cdots ,\hat{\sigma}_j(t),\cdots ,\sigma
_N(t)\right) $, in general, depends on the momentary values of the
neighboring spins as well as on the influence of the heat bath. For this
reason statistical correlations exist between different spins. Therefore, it
is necessary to deal with the entire $N$-spin system as a unit. The
evolution of spin functions describing system form a Markov process of $N$
discrete or continuous random variables with a continuous time variable as
argument.

We introduce a probability distribution function $p(\sigma _1,\cdots ,\sigma
_N,t)$, which denotes the probability of spin system being in the state $%
(\sigma _1,\cdots ,\sigma _N)$ at time $t$. Let $W_j(\sigma _j\rightarrow 
\hat{\sigma}_j)$ be the transition probability per unit time that the $j$th
spin transits from one value $\sigma _j$ to another possible value $\hat{%
\sigma}_j$, while the others remain fixed. Then, on the supposition of
single-spin transition, we may write the time derivative of the function $%
P(\sigma _1,\cdots ,\sigma _N,t)$ as 
\begin{equation}
\frac d{dt}P(\{\sigma \},t)=\sum_j\sum_{\hat{\sigma}_j}\left[ -W_j(\sigma
_j\rightarrow \hat{\sigma}_j)p(\{\sigma \},t)+W_j(\hat{\sigma}_j\rightarrow
\sigma _j)p(\{\sigma _{i\neq j}\},\hat{\sigma}_j,t)\right] .  \label{1}
\end{equation}
This is a probability equation, in which the first term in the hand-right
side of Eq.(\ref{1}) denotes the decrease of the probability distribution
function $P(\{\sigma \},t)$ per unit time due to the transition of the spin
state from the initial value $\sigma _{j\text{ ,}}(j=1,2,\cdots ,N)$ to
various possible final values $\hat{\sigma}_j$, and the second term denotes
the increase of the probability distribution function $P(\{\sigma \},t)$ per
unit time due to the transition of the spin state from the various possible
initial values $\hat{\sigma}_j,(j=1,2,\cdots ,N)$ to final value $\sigma $$%
_j $. We shall refer to the Eq.(\ref{1}) as the master equation since its
solution would contain the most complete description of the system available.

It is the most crucial step, obviously, that the transition probability must
be determined before the master equation can be solved. Then, how to
determine the transition probability? For this problem, Glauber' theory
leaves some leeway to select. However, inappropriate selection will probably
make the problem difficult to solve. So, we hope to find out a more definite
expression to apply the Glauber' theory to arbitrary-spin systems. Now we
consider it from both mathematics and physics aspects. In mathematics,
generally speaking, the probability must be ergodic and positive definite,
and can be normalized; in physics, we often request that an equilibrium
thermodynamic system satisfies the detailed balance condition. Based on
these considerations, we can restrict the spin transition probability $%
W_j\left( \sigma _j\rightarrow \hat{\not{\sigma}}_j\right) $ to satisfy the
following conditions so that ensure the system in thermodynamic equilibrium
state finally:

For the set $(S,\hat{S})$ composed of a subset $S$ and its dual subset $\hat{%
S}$ in phase space, existing $\sigma _j$ belonging to $S$ and $\hat{\sigma}%
_j $ belonging to $\hat{S}$ , we have

(1) ergodicity 
\begin{equation}
\forall \sigma _j,\hat{\sigma}_j:\ W_j(\sigma _j\rightarrow \hat{\sigma}%
_j)\neq 0;  \label{2}
\end{equation}
\qquad \qquad \qquad \qquad

(2) positivity 
\begin{equation}
\forall \sigma _j,\hat{\sigma}_j:\ W_j(\sigma _j\rightarrow \hat{\sigma}%
_j)\geq 0;  \label{3}
\end{equation}

(3) normalization 
\begin{equation}
\forall \sigma _j:\ \sum_{\hat{\sigma}_j}W_j(\sigma _j\rightarrow \hat{\sigma%
}_j)=1;  \label{4}
\end{equation}

(4) detailed balance 
\begin{equation}
\forall \sigma _j,\hat{\sigma}_j:\ \frac{W_j(\sigma _j\rightarrow \hat{\sigma%
}_j)}{W_j(\hat{\sigma}_j\rightarrow \sigma _j)}=\frac{P_{eq}(\sigma
_1,\cdots ,\hat{\sigma}_j,\cdots ,\sigma _N)}{P_{eq}(\sigma _1,\cdots
,\sigma _j,\cdots ,\sigma _N)},  \label{5}
\end{equation}
in which 
\[
P_{eq}=\frac 1Z\exp [-\beta {\cal H}(\{\sigma \})],\quad Z=\sum_{\{\sigma
\}}\exp [-\beta {\cal H}(\{\sigma \})], 
\]
where $P_{eq}$ is the equilibrium Boltzmann distribution function, $Z$ is
the partition function, ${\cal H}(\{\sigma \})$ is the system Hamiltonian.

Although the spin transition probabilities are not determined uniquely by
above restriction conditions, there are less room to choose them.
Furthermore, considering the fact that the transition of the individual spin
depends merely on the momentary values of the neighboring spins as well as
on the influence of the heat bath, we can suppose that the transition
probability from $\sigma _{j\text{ }}$ to $\hat{\sigma}_j$ depend only on
the heat Boltzmann factor of the neighboring spins, i.e. 
\[
W_i(\sigma _i\rightarrow \hat{\sigma}_i)\propto \exp \left[ -\beta {\cal H}%
_i\left( \hat{\sigma}_i,\sum_{<ij>}\sigma _j\right) \right] , 
\]
or 
\begin{equation}
W_i(\sigma _i\rightarrow \hat{\sigma}_i)=\frac 1{Q_i}\exp \left[ -\beta 
{\cal H}_i\left( \hat{\sigma}_i,\sum_{<ij>}\sigma _j\right) \right] ,
\label{6}
\end{equation}
where $\sum_{<ij>}$ means that the summation for $j$ is only related to the
neighboring values of $i$. By means of the normalized condition (\ref{4}),
the normalized factor $Q_i$ can be determined as 
\begin{equation}
Q_i=\sum_{\hat{\sigma}_i}\exp \left[ -\beta {\cal H}_i\left( \hat{\sigma}%
_i,\sum_{<ij>}\sigma _j\right) \right] .  \label{7}
\end{equation}
Obviously, $Q_i$ is independent of $\sigma _i$ and is related to the
temperature and neighboring spins.

Compared with Glauber' expression[1], Eq.(\ref{6}) is a normalized version
of transition probability. As we know, the constant $\alpha $ in Glauber'
expression is a free constant determined by the time scale. Actuaclly, our
expression is only a definite selection for constant $\alpha $ by extra
restriction conditions and physical considerations.

Usually, we are interested in local magnetization and the spin-pair
correlation, they are defined as follows[1] 
\begin{equation}
q_k(t)=<\sigma _k(t)>=\sum_{\{\sigma \}}\sigma _kP(\{\sigma \},t),  \label{8}
\end{equation}
\begin{equation}
\gamma _{kl}(t)=<\sigma _k(t)\sigma _l(t)>=\sum_{\{\sigma \}}\sigma _k\sigma
_lP(\{\sigma \},t).  \label{9}
\end{equation}
According to the definitions (\ref{8})-(\ref{9}) and the master equation (%
\ref{1}), and using the normalized condition (\ref{4}), time-evolving
equations of $q_k(t)$ and $\gamma _{kl}(t)$ can be derived ( see Appendix 
\ref{app-a}) 
\begin{equation}
\frac d{dt}q_k(t)=-q_k(t)+\sum_{\{\sigma \}}\left( \sum_{\hat{\sigma}_k}\hat{%
\sigma}_kW_k(\sigma _k\rightarrow \hat{\sigma}_k)\right) P(\{\sigma \},t),
\label{10}
\end{equation}
\begin{eqnarray}
\frac d{dt}\gamma _{kl}(t) &=&-2\gamma _{kl}(t)+\sum_{\{\sigma \}}\left[
\sigma _k\left( \sum_{\hat{\sigma}_l}\hat{\sigma}_lW_l(\sigma _l\rightarrow 
\hat{\sigma}_l)\right) \right.  \nonumber \\
&&\left. +\sigma _l\left( \sum_{\hat{\sigma}_k}\hat{\sigma}_kW_k(\sigma
_k\rightarrow \hat{\sigma}_k)\right) \right] P(\{\sigma \},t).  \label{11}
\end{eqnarray}
Similarly, time-evolving equation of equal-time multi-spin correlation
function can be further derived 
\begin{eqnarray}
\frac d{dt} &<&\sigma _{i_1}(t)\sigma _{i2}(t)\cdots \sigma
_{i_n}(t)>=-n<\sigma _{i_1}\sigma _{i_2\cdots }\sigma _{i_n}>  \nonumber \\
&&+\sum_{\{\sigma \}}\left\{ \sum_{k=1}^n\left[ \left( \prod_{j(\neq
k)=1}^n\sigma _{i_j}\right) \left( \sum_{\hat{\sigma}_{i_k}}\hat{\sigma}%
_{i_k}W_{i_k}(\sigma _{i_k}\rightarrow \hat{\sigma}_{i_k})\right) \right]
\right\} P(\{\sigma \},t).  \label{12}
\end{eqnarray}
Eqs.(\ref{1}), (\ref{4})--(\ref{7}) and (\ref{10})--(\ref{12}) are the basic
formulas of the single-spin transition-type critical dynamics which suits
one-dimensional arbitrary-spin systems.

All of these formulas can readily extended to spin systems on square lattice
and cubic lattice. The only corrections are changing subscript $i$ into $ij$
and $ijk$ respectively.

\section{Exact results}

\label{Sec.3}

In this section, three examples of the application are given, including
one-, two-, and three-dimensional kinetic Gaussian model.

Now we treat the kinetic Gaussian model. First of all we will introduce
Gaussian model, then we will exactly solve the evolution of the local
magnetization and equal-time spin-pair correlation function, and we will
obtain the dynamical exponent $z$. Here, we will only give the solving
process of three-dimensional case in detail.

The Gaussian model, proposed by T.H. Berlin and M.Kac \cite{20} at first in
order to make Ising model more tractable, is an continuous-spin model.
Comparing with the Ising model, besides having the same Hamiltonian form
(three-dimensional case) 
\begin{equation}
-\beta {\cal H}{\bf (}\sigma _{ijk})=k\sum_{i,j,k=1}^N\sum_w\sigma
_{ijk}\left( \sigma _{i+w,j,k}+\sigma _{i,j+w,k}+\sigma _{i,j,k+w}\right) ,
\label{3.1}
\end{equation}
where $\sum_w$ meas summation over near neighbors, there are two extension:
First, the spin $\sigma _{ijk}$ can take any real value between ($-\infty
,+\infty $). Secondly, to prevent from all spins tend to infinity, a
probability of finding a given spin between $\sigma _{ijk}$ and $\sigma
_{ijk}+d\sigma _{ijk}$ is assumed to be the Gaussian-type distribution 
\begin{equation}
f(\sigma _{ijk})d\sigma _{ijk}=\sqrt{\frac b{2\pi }}\exp \left[ -\frac b2%
\sigma _{ijk}^2\right] d\sigma _{ijk},  \label{3.2}
\end{equation}
where $b$ is a distribution constant independent of temperature. Although
being an extension of Ising model, Gaussian model shows much difference from
Ising model in the properties of phase transition. In equilibrium case, on
translational invariant lattices the Gaussian model was exactly solvable,
and later as a starting point to study the unsolvable models it was also
investigated with mean field theory and momentum-space renormalization-group
method[21-22].Recently the Gaussian model on fractal lattices was studied by
Li and Yang[23]. However, the critical dynamic problem of continuous-spin
model have never been investigated so far.

We now proceed to treat the isotropic kinetic Gaussian model on the cube
lattice. The system Hamiltonian and the spin distribution probability are (%
\ref{3.1}) and (\ref{3.2}) respectively. In this case the spin transition
probability can be expressed as

\begin{equation}
W_{ijk}(\sigma _{ijk}\rightarrow \hat{\sigma}_{ijk})=\frac 1{Q_{ijk}}\exp
\left[ k\hat{\sigma}_{ijk}\sum_w\left( \sigma _{i+w,j,k}+\sigma
_{i,j+w,k}+\sigma _{i,j,k+w}\right) \right] .  \label{3.3}
\end{equation}
Because the spin take continuous value, the summation for spin value turns
into the integration 
\begin{equation}
\sum_\sigma \rightarrow \int_{-\infty }^\infty f(\sigma )d\sigma ,
\label{3.4}
\end{equation}
then the normalized factor $Q_{ijk}$ can be determined as 
\begin{eqnarray}
Q_{ijk} &=&\int_{-\infty }^\infty d\hat{\sigma}_{ijk}f(\hat{\sigma}%
_{ijk})\exp \left[ k\hat{\sigma}_{ijk}\sum_w\left( \sigma _{i+w,j,k}+\sigma
_{i,j+w,k}+\sigma _{i,j,k+w}\right) \right]  \nonumber \\
&=&\exp \left\{ \frac{k^2}{2b}\left[ \sum_w\left( \sigma _{i+w,j,k}+\sigma
_{i,j+w,k}+\sigma _{i,j,k+w}\right) \right] ^2\right\} ,  \label{3.5}
\end{eqnarray}
and the another useful combination formula can also be obtained 
\begin{eqnarray}
\sum_{\hat{\sigma}_{ijk}}\hat{\sigma}_{ijk}W_{ijk}(\sigma _{ijk}
&\rightarrow &\hat{\sigma}_{ijk})=\int_{-\infty }^\infty \hat{\sigma}%
_{ijk}W_{ijk}(\sigma _{ijk}\rightarrow \hat{\sigma}_{ijk})f(\hat{\sigma}%
_{ijk})d\hat{\sigma}_{ijk}  \nonumber \\
&=&\frac kb\sum_w\left( \sigma _{i+w,j,k}+\sigma _{i,j+w,k}+\sigma
_{i,j,k+w}\right) .  \label{3.6}
\end{eqnarray}
Substituting (\ref{3.6}) into the following time-evolving equations of the
local magnetization and the equal-time spin-pair correlation function 
\begin{equation}
\frac d{dt}q_{ijk}(t)=-q_{ijk}(t)+\sum_{\{\sigma \}}\left( \sum_{\hat{\sigma}%
_{ijk}}\hat{\sigma}_{ijk}w_{ijk}(\sigma _{ijk}\rightarrow \hat{\sigma}%
_{ijk})\right) P(\{\sigma \},t),  \label{3.7}
\end{equation}
\begin{eqnarray}
\frac d{dt}\gamma _{ijk;i^{\prime }j^{\prime }k^{\prime }}(t) &=&-2\gamma
_{ikj;i^{\prime }j^{\prime }k^{\prime }}(t)+\sum_{\{\sigma \}}\left[ \sigma
_{ijk}\left( \sum_{\hat{\sigma}_{i^{\prime }j^{\prime }k^{\prime }}}\hat{%
\sigma}_{i^{\prime }j^{\prime }k^{\prime }}w_{i^{\prime }j^{\prime
}k^{\prime }}(\sigma _{i^{\prime }j^{\prime }k^{\prime }}\rightarrow \hat{%
\sigma}_{i^{\prime }j^{\prime }k^{\prime }})\right) \right.  \nonumber \\
&&\left. +\sigma _{i^{\prime }j^{\prime }k^{\prime }}\left( \sum_{\hat{\sigma%
}_{ijk}}\hat{\sigma}_{ijk}w_{ijk}(\sigma _{ijk}\rightarrow \hat{\sigma}%
_{ijk})\right) \right] \ P(\{\sigma \},t),  \label{3.8}
\end{eqnarray}
we get 
\begin{equation}
\frac d{dt}q_{ijk}(t)=-q_{ijk}(t)+\frac kb\sum_w\left(
q_{i+w,j,k}+q_{i,j+w,k}+q_{i,j,k+w}\right) ,  \label{3.9}
\end{equation}
\begin{eqnarray}
\frac d{dt}\gamma _{ijk;i^{\prime }j^{\prime }k^{\prime }}(t) &=&-2\gamma
_{ijk;i^{\prime }j^{\prime }k^{\prime }}(t)+\frac kb\sum_w\left( \gamma
_{ijk;i^{\prime }+w,j^{\prime },k^{\prime }}(t)+\gamma _{ijk;i^{\prime
},j^{\prime }+w,k^{\prime }}(t)\right.  \nonumber \\
&&+\left. \gamma _{ijk;i^{\prime },j^{\prime },k^{\prime }+w}(t)+\gamma
_{i+w,j,k;i^{\prime }j^{\prime }k^{\prime }}(t)+\gamma _{i,j+w,k;i^{\prime
}j^{\prime }k^{\prime }}(t)+\gamma _{i,j,k+w;i^{\prime }j^{\prime }k^{\prime
}}(t)\right) .  \label{3.10}
\end{eqnarray}

In order to solve Eqs.(\ref{3.9}) and (\ref{3.10}) in the nearest-neighbor
interaction case $\left( w=\pm 1\right) $, we introduce two generating
functions[1]: 
\begin{equation}
F_1(\lambda _1,\lambda _2,\lambda _3,t)=\sum_{i,j,k=-\infty }^\infty \lambda
_1^i\lambda _2^j\lambda _3^kq_{ijk}(t),  \label{3.11}
\end{equation}
and 
\begin{equation}
F_2(\lambda _1,\cdots ,\lambda _6,t)=\sum_{i,j,k;i^{\prime },j^{\prime
},k^{\prime }=-\infty }^\infty \lambda _1^i\lambda _2^j\lambda _3^k\lambda
_4^{i^{\prime }}\lambda _5^{j^{\prime }}\lambda _6^{k^{\prime }}\gamma
_{ijk;i^{\prime }j^{\prime }k^{\prime }}(t),  \label{3.12}
\end{equation}
then the Eqs.(\ref{3.9}) and (\ref{3.10}) turn into the following equations
with resect to $F_1$ and $F_2$, respectively, 
\begin{equation}
\frac d{dt}F_1(\lambda _1,\lambda _2,\lambda _3,t)=\left[ -1+\frac kb%
\sum_{i=1}^3\left( \lambda _i+\lambda _i^{-1}\right) \right] F_1(\lambda
_1,\lambda _2,\lambda _3,t),  \label{3.13}
\end{equation}
\begin{equation}
\frac d{dt}F_2(\lambda _1,\cdots ,\lambda _6,t)=\left[ -2+\frac kb%
\sum_{i=1}^6\left( \lambda _i+\lambda _i^{-1}\right) \right] F_2(\lambda
_1,\cdots ,\lambda _6,t).  \label{3.14}
\end{equation}
Solving Eqs.(\ref{3.13}) and (\ref{3.14}), we get 
\begin{equation}
F_1(\lambda _1,\lambda _2,\lambda _3,t)=F_1(\lambda _1,\lambda _2,\lambda
_3,0)e^{-t}\exp \left[ \frac kb\sum_{i=1}^3\left( \lambda _i+\lambda
_i^{-1}\right) t\right] ,  \label{3.15}
\end{equation}
\begin{equation}
F_2(\lambda _1,\cdots ,\lambda _6,t)=F_2(\lambda _1,\cdots ,\lambda
_6,0)e^{-2t}\exp \left[ \frac kb\sum_{i=1}^6\left( \lambda _i+\lambda
_i^{-1}\right) t\right] .  \label{3.16}
\end{equation}
In terms of a generating function of the Bessel functions of imaginary
argument 
\begin{equation}
e^{x(\lambda +\lambda ^{-1})/2}=\sum_{v=-\infty }^\infty \lambda ^vI_v(x),
\label{3.17}
\end{equation}
we obtain immediately the following exact solutions 
\begin{equation}
q_{ijk}(t)=e^{-t}\sum_{n,m,l=-\infty }^\infty q_{nml}(0)I_{i-n}\left( \frac{%
2k}bt\right) I_{j-m}\left( \frac{2k}bt\right) I_{k-l}\left( \frac{2k}b%
t\right) ,  \label{3.18}
\end{equation}
\begin{eqnarray}
\gamma _{ijk;i^{\prime }j^{\prime }k^{\prime }}(t)
&=&e^{-2t}\sum_{n,m,l;n^{\prime },m^{\prime },l^{\prime }=-\infty }^\infty
\gamma _{nml;n^{\prime }m^{\prime }l^{\prime }}(0)  \nonumber \\
&&\times I_{i-n}\left( \frac{2k}bt\right) I_{j-m}\left( \frac{2k}bt\right)
I_{k-l}\left( \frac{2k}bt\right) I_{i^{\prime }-n^{\prime }}\left( \frac{2k}b%
t\right) I_{j^{\prime }-m^{\prime }}\left( \frac{2k}bt\right) I_{k^{\prime
}-l^{\prime }}\left( \frac{2k}bt\right) ,  \nonumber \\
&&  \label{3.19}
\end{eqnarray}
where $q_{nml}(0)$ and $\gamma _{nml;n^{\prime }m^{\prime }l^{\prime }}(0)$,
respectively, correspond to their initial values.

By using of the asymptotic expansion expression of the first-kind imaginary
argument Bessel function 
\begin{eqnarray}
I_v(x) &=&\frac{e^x}{\sqrt{2\pi x}}\sum_{n=0}^\infty \frac{(-)^n(v,n)}{(2x)^n%
}+\frac{e^{-x+(v+\frac 12)\pi i}}{\sqrt{2\pi x}}\sum_{n=0}^\infty \frac{(v,n)%
}{(2x)^n},(-\pi /2<\arg x<3\pi /2),\left| x\right| \rightarrow \infty , 
\nonumber  \label{3.35} \\
&&  \label{3.20}
\end{eqnarray}
where 
\[
(v,n)=\frac{\Gamma \left( \frac 12+v+n\right) }{n!\Gamma \left( \frac 12%
+v-n\right) }, 
\]
we can obtain the long-time asymptotic behavior of the local magnetization 
\begin{equation}
q_{ijk}(t)\sim (\frac{2k}bt)^{-3/2}e^{-(1-6k/b)t}\sum_{n,m,l=-\infty
}^\infty q_{nml}(0)\sim \frac 1{t^{3/2}}e^{-t/\tau },  \label{3.21}
\end{equation}
\begin{equation}
\tau =\frac 1{1-6k/b},  \label{3.22}
\end{equation}
where $\tau $ is the relaxation time of the system. We know that the
critical point of the Gaussian model is $k_c=J/k_\beta T_c=b/2d$, where $d$
is the spatial dimension, and the correlation length critical exponent is $%
\nu =1/2$ [22]. So, by means of the following dynamical scaling hypotheses 
\begin{equation}
\xi \sim \left| T-T_c\right| ^{-\nu },  \label{3.23}
\end{equation}
\begin{equation}
\tau \sim \xi ^z,  \label{3.24}
\end{equation}
the dynamic critical exponent $z$ of the 3D kinetic Gaussian model can be
obtained 
\begin{equation}
z=2.  \label{3.25}
\end{equation}

With the same treatment, we can easily solve one- and two-dimensional
kinetic Gaussian model. Ignoring the process of solution, we give only the
following exact results:

(1). 1D case

\begin{equation}
q_k(t)=e^{-t}\sum_{m=-\infty }^\infty q_m(0)I_{k-m}(\frac{2k}bt),
\label{3.26}
\end{equation}
\begin{equation}
\gamma _{kl}(t)=e^{-2t}\sum_{n,m=-\infty }^\infty \gamma _{nm}(0)I_{k-n}(%
\frac{2k}bt)I_{l-m}(\frac{2k}bt),  \label{3.27}
\end{equation}
\begin{equation}
\tau =\frac 1{1-2k/b},  \label{3.28}
\end{equation}
\begin{equation}
z=2.  \label{3.29}
\end{equation}

(2). 2D case

\begin{equation}
q_{nm}(t)=e^{-t}\sum_{k,l=-\infty }^\infty q_{k,l}(0)I_{n-k}\left( \frac{2k}b%
t\right) I_{m-l}\left( \frac{2k}bt\right) ,  \label{3.30}
\end{equation}
\begin{equation}
\gamma _{mn;m^{\prime }n^{\prime }}(t)=e^{-2t}\sum_{i,j;i^{\prime
},j^{\prime }=-\infty }^\infty \gamma _{ij;i^{\prime }j^{\prime
}}(0)I_{m-i}\left( \frac{2k}bt\right) I_{n-j}\left( \frac{2k}bt\right)
I_{m^{\prime }-i^{\prime }}\left( \frac{2k}bt\right) I_{n^{\prime
}-j^{\prime }}\left( \frac{2k}bt\right)  \label{3.31}
\end{equation}
\begin{equation}
\tau =\frac 1{1-4k/b},  \label{3.32}
\end{equation}
\begin{equation}
z=2.  \label{3.33}
\end{equation}

\section{Conclusion}

\label{Sec.4}

In this paper, we have suggested a normalized transition probability to
solve the time evolution equations of the local magnetization and spin-pair
correlation function. Our treatment can in part remove the arbitrariness in
Glauber dynamical theory, and makes exactly solve the time evolution
equation possible.

Based on our treatment, we have re-investigated one-dimensional kinetic
Ising model, resulting in same dynamical critical exponent $z$ as
Glauber[1]. We have also exactly solved the kinetic Gaussian model, and
given the detail of solving process of three-dimensional case. We have in
surprise found that the dynamical critical exponents have same value
independent of spatial dimension, which shows the dynamical behavior has
super-universality in contrast with static behavior. In fact, in equilibrium
phase transition the critical exponents are strongly dependent on
dimensionality.

\acknowledgments 

The work was supported by the National Basic Research Project ``Nonlinear
Science'' and the National Natural Science Foundation of China.

\appendix 

\section{Proofs of Eqs. (10)--(12)}

\label{app-a}

According to the definition (\ref{8}) and using the master equation (\ref{1}%
), we have 
\begin{eqnarray}
\frac d{dt}q_k(t) &=&\frac d{dt}\sum_{\{\sigma \}}\sigma _kP(\{\sigma \},t) 
\nonumber \\
&=&\sum_{\{\sigma \}}\sum_i\sum_{\hat{\sigma}_i}\left[ -\sigma _kw_i(\sigma
_i\rightarrow \hat{\sigma}_i)P(\{\sigma \},t)+\sigma _kw_i(\hat{\sigma}%
_i\rightarrow \sigma _i)P(\{\sigma _{j\neq i}\},\hat{\sigma}_i,t)\right] 
\nonumber \\
&=&\sum_{\{\sigma \}}\sigma _k\sum_{i(i\neq k)}\sum_{\hat{\sigma}_i}\left[
-w_i(\sigma _i\rightarrow \hat{\sigma}_i)P(\{\sigma \},t)+w_i(\hat{\sigma}%
_i\rightarrow \sigma _i)P(\{\sigma _{j\neq i}\},\hat{\sigma}_i,t)\right] 
\nonumber \\
&&+\sum_{\{\sigma \}}\sum_{\hat{\sigma}_k}\left[ -\sigma _kw_k(\sigma
_k\rightarrow \hat{\sigma}_k)P(\{\sigma \},t)+\sigma _kw_k(\hat{\sigma}%
_k\rightarrow \sigma _k)P(\{\sigma _{j\neq k}\},\hat{\sigma}_k,t)\right]
\label{A.1}
\end{eqnarray}
Looking at the first term ($i\neq k$) in the latter of the last equality
sign of (\ref{A.1}), 
\begin{eqnarray}
\mbox{$(i\neq k)$ term} &=&\sum_{\{\sigma \}}\sigma _k\sum_{i(i\neq k)}\sum_{%
\hat{\sigma}_i}[-w_i(\sigma _i\rightarrow \hat{\sigma}_i)P(\{\sigma
\},t)+w_i(\hat{\sigma}_i\rightarrow \sigma _i)P(\{\sigma _{j\neq i}\},\hat{%
\sigma}_i,t)]  \nonumber \\
&=&\sum_{\{\sigma _{j\neq i}\}}\sigma _k\sum_{i(i\neq k)}[-\sum_{\sigma _i,%
\hat{\sigma}_i}w_i(\sigma _i\rightarrow \hat{\sigma}_i)P(\{\sigma
\},t)+\sum_{\sigma _i,\hat{\sigma}_i}w_i(\hat{\sigma}_i\rightarrow \sigma
_i)P(\{\sigma _{j\neq i}\},\hat{\sigma}_i,t)]  \nonumber \\
&&  \label{A.2}
\end{eqnarray}
it is easy to see that this term equals to zero, as long as $\hat{\sigma}_i$
exchange with $\sigma _i$ before doing sum for $\hat{\sigma}_i$ and $\sigma
_i$. So the surplus term of (\ref{A.1}) is only the last term ($i=k$): 
\begin{eqnarray*}
\frac d{dt}q_k(t) &=&\frac d{dt}\sum_{\{\sigma \}}\sigma _kP(\{\sigma \},t)
\\
&=&\sum_{\{\sigma \}}\sum_{\hat{\sigma}_k}\left[ -\sigma _kw_k(\sigma
_k\rightarrow \hat{\sigma}_k)P(\{\sigma \},t)+\sigma _kw_k(\hat{\sigma}%
_k\rightarrow \sigma _k)P(\{\sigma _{j\neq k}\},\hat{\sigma}_k,t)\right] \\
&=&-\sum_{\{\sigma \}}\sigma _k\left( \sum_{\hat{\sigma}_k}w_k(\sigma
_k\rightarrow \hat{\sigma}_k)\right) P(\{\sigma \},t)+\sum_{\sigma _1\cdots
\sigma _k\hat{\sigma}_k\cdots \sigma _N}\sigma _kw_k(\hat{\sigma}%
_k\rightarrow \sigma _k)P(\{\sigma _{j\neq k}\},\hat{\sigma}_k,t) \\
&=&-\sum_{\{\sigma \}}\sigma _kP(\{\sigma \},t)+\sum_{\sigma _1\cdots \hat{%
\sigma}_k\sigma _k\cdots \sigma _N}\hat{\sigma}_kw_k(\sigma _k\rightarrow 
\hat{\sigma}_k)P(\{\sigma _{j\neq k}\},\sigma _k,t) \\
&=&-q_k(t)+\sum_{\{\sigma \}}\left( \sum_{\hat{\sigma}_k}\hat{\sigma}%
_kw_k(\sigma _k\rightarrow \hat{\sigma}_k)\right) P(\{\sigma \},t),
\end{eqnarray*}
in which, the normalized condition and the technic of exchange of $\hat{%
\sigma}_i$ for $\sigma _i$ were used. Hitherto, Eq.(10) have been proven
exactly. As for the proof of the Eqs.(11) and (12), it is easily accessible
via the same method, so without the necessity for further proof.


\begin{references}
\bibitem{1}  R. J. Glauber, J. Math. Phys. {\bf 4}, 294 (1963).

\bibitem{2}  K. Kawasaki, Phys. Rev. {\bf 145}, 224 (1965).

\bibitem{3}  G. Forgacs, S. T. Chui and H. L. Frisch, Phys. Rev. B {\bf 22},
415 (1980).

\bibitem{4}  E. J. Lage, J. Phys. A: Math. Gen. {\bf 18,} 2289 (1985).

\bibitem{5}  E. J. Lage, J. Phys. A: Math. Gen. {\bf 18,} 2411 (1985).

\bibitem{6}  E. J. Lage, Phys. Lett. A {\bf 127}, 9 (1988).

\bibitem{7}  Y. Achiam, Phys. Rev. B {\bf 31}, 4732 (1985).

\bibitem{8}  Y. Achiam, Phys. Rev. B {\bf 32}, 1796 (1985).

\bibitem{9}  Y. Achiam, Phys. Rev. B {\bf 33}, 7762 (1986).

\bibitem{10}  Jian Zhou and Z. R. Yang, Phys. Rev. B {\bf 39}, 9423 (1989).

\bibitem{11}  M. D. Lacasse, J. Vinals and M.G rant, Phys. Rev. B {\bf 47},
5646 (1993).

\bibitem{12}  B. C. S. Grandi and W. Figueiredo, Phys. Rev. E {\bf 54}, 4722
(1996).

\bibitem{13}  J. Rogiers and J. O. Indeleu, Phys. Rev. B {\bf 41}, 6998
(1990).

\bibitem{14}  J. Wang, Phys. Rev. B {\bf 47}, 896 (1993).

\bibitem{15}  Y. Achiam and J. M. Kosterlitz, Phys. Rev. Lett. {\bf 41}, 128
(1978).

\bibitem{16}  G. F. Mazenko and O. T. Valls, Phys. Rev. B{\bf \ 24}, 1419
(1981).

\bibitem{17}  P. C. Hoohenberg and B. I. Halperin, Rev. Mod. Phys. {\bf 49},
435 (1977).

\bibitem{18}  Z. R. Yang, Phys. Rev. B {\bf 46}, 11578 (1992).

\bibitem{19}  Peiqing Tong, Phys. Rev. E {\bf 56}, 1371 (1997).

\bibitem{20}  T. H. Berlin and M. Kac, Phys. Rev. {\bf 86}, 821 (1952).

\bibitem{21}  J. J. Binney, N. J. Dowrick, A. J. Fisher and M. E. J. newman,
The Theory of Critical Phenomena, Clarendon Press, Oxford 1992.

\bibitem{22}  L. E. Riechl, A Modern course in Statistical Physics,
University of Texas Press, Austin, 1980.

\bibitem{23}  Song Li and Z. R. Yang, Phys. Rev. E {\bf 55}, 6656 (1997).
\end{references}
\end{document}